\documentclass[sigconf, nonacm]{acmart}

\usepackage{amsmath,amsfonts}
\usepackage{algorithmic}
\usepackage{graphicx}
\usepackage{textcomp}
\usepackage{xcolor}
\usepackage{booktabs}
\usepackage{xcolor}
\usepackage{pgfplots}
\usepackage{subcaption}
\usepackage{colortbl}
\usepackage{braket}
\usepackage{epsfig,latexsym,graphics}
\usepackage[font={small}]{caption}
\usepackage{setspace}
\usepackage{enumitem}
\pgfplotsset{compat=1.16}
\usepackage{cmd}
\usepackage{svg}
\usepackage[spacesep,definevectors]{easyvector}

\usepackage{balance}

\begin{document}

\title{ESG: Elastic Graphs for Range-Filtering\\
Approximate $k$-Nearest Neighbor Search}

\author{Mingyu Yang$^{1,2}$, Wentao Li$^{3}$, Zhitao Shen$^{4}$, Chuan Xiao$^{5}$, Wei Wang$^{1,2}$}
\affiliation{%
\institution{The Hong Kong University of Science and Technology (Guangzhou)\country{China}$^{1}$}
\institution{The Hong Kong University of Science and Technology\country{China}$^{2}$}
\institution{University of Leicester\country{UK}$^{3}$,Ant Group\country{China}$^{4}$,Osaka University\country{Japan}$^{5}$}
}
\email{myang250@connect.hkust-gz.edu.cn, wl226@leicester.ac.uk}
\email{zhitao.szt@antgroup.com,chuanx@ist.osaka-u.ac.jp, weiwcs@ust.hk}


\begin{abstract}
Range-filtering approximate $k$-nearest neighbor (RFAKNN) search takes as input a vector and a numeric value, returning $k$ points from a database of $N$ high-dimensional points. 
The returned points must satisfy two criteria: their numeric values must lie within the specified query range, and they must be approximately the $k$ nearest points to the query vector.
A straightforward approach to processing RFAKNN queries is to filter out points outside the query range and then perform a search on the remaining in-range points. 
While indexing can speed up this process, ensuring correctness would require indexing all possible $O(N^2)$ query ranges, which is infeasible due to severe storage constraints.
Lossy compression techniques have been explored to create compact indexes for these ranges, but they often compromise query accuracy because of the inherent loss of information.
To address these challenges, existing methods index only a subset of ranges, organized within a segment tree. 
At query time, the query range is decomposed into multiple subranges stored in the segment tree. 
Although these methods improve accuracy, they introduce significant query overhead since processing a single query requires combining indexes from $O(\log N)$ subranges.
To strike a better balance between query accuracy and efficiency, we propose novel methods that relax the strict requirement for subranges to \textit{exactly} match the query range. 
This elastic relaxation is based on a theoretical insight: allowing the controlled inclusion of out-of-range points during the search does not compromise the bounded complexity of the search process. 
Building on this insight, we prove that our methods reduce the number of required subranges to at most \textit{two}, eliminating the $O(\log N)$ query overhead inherent in existing methods.
Extensive experiments on real-world datasets demonstrate that our proposed methods outperform state-of-the-art approaches, achieving performance improvements of 1.5x to 6x while maintaining high accuracy.
\end{abstract}

\maketitle

\section{Introduction}
The problem of $k$-nearest neighbor (KNN) search in high-dimensional spaces has received increasing attention, especially with recent advancements in deep learning and large language models~\cite{LLM-RAG-NIPS-2020}. 
Due to the curse of dimensionality~\cite{Curse-of-dim-1998}, finding the exact $k$ nearest neighbors often becomes computationally demanding. 
As a result, the variant known as \textbf{approximate $k$-nearest neighbor (AKNN)} search has been extensively studied. 
Numerous methods have been proposed for AKNN search~\cite{LSH-1999,LSH-Pstable-2004,QALSH-2015-VLDB,PMLSH-bolong-2020,SRS-yifang-2014,C2LSH-2012-SIGMOD,Binary-LSH-NIPS-2009,DBLP:journals/pami/OPQGeHK014,PQ-fast-scan-VLDB-2015,LOPQ-2014-CVPR,DeltaPQ-VLDB-2020,Rabitq-SIGMOD-2024,Cover-Tree-2006-ICML,bolt-KDD-2017,BLISS-KDD-2022,IP-PQ-2016-guoruiqi,RoutingPQ-2024-ICDE,DBLP:journals/pami/PQJegouDS11,IMI:DBLP:journals/pami/BabenkoL15}, among which graph-based approaches~\cite{FANNG:harwood2016fanng,HVS:DBLP:journals/pvldb/LuKXI21,tMRNG:journals/pacmmod/PengCCYX23,DBLP:journals/pvldb/NSGFuXWC19,DBLP:journals/pami/SSGFuWC22,Diskann-NIPS-2019} have emerged as particularly effective.
Graph-based methods, such as $\HNSW$~\cite{DBLP:journals/pami/HNSWMalkovY20}, build graphs as the index for AKNN search and demonstrate superior performance compared to alternative methods.

Driven by the demands of applications, \textbf{Range-Filtering Approximate $k$-Nearest Neighbors (RFAKNN)} search extends the traditional AKNN search problem by incorporating additional numerical constraints. 
In such scenarios, data points typically consist of two components: a vector and an associated numerical attribute.
Formally, given a database $\mathcal{D}$ of $N$ data points, where each point $v_i$ is associated with a numerical value\footnote{We apply the re-ranking strategy in~\cite{SeRF-SIGMOD-2024} to map the attribute value of a point $v_i$ to its position $i$ in $\mathcal{D}$. Thus, both the cardinality of $\mathcal{D}$ and the range are $N = |\mathcal{D}|$.} $i$, an RFAKNN query takes as input a vector $q$ and a range $[l_q, r_q]$, and returns $k$ points as the result. 
These points should satisfy two conditions: their numerical values must lie within the query range $[l_q, r_q]$, and they must be approximately the $k$ closest points to the query vector $q$. 
By introducing a range as a filter to exclude data points outside the specified range, RFAKNN enables more targeted and meaningful nearest-neighbor retrieval. For example, in e-commerce applications, each product typically has a price attribute. 
By specifying a price range during an AKNN search, users can filter out products beyond their budget, improving the relevance of search results. 
Similarly, RFAKNN has broad applicability in Retrieval-Augmented Generation (RAG)~\cite{Window-Filter-ICML-2024}, vehicle search~\cite{SeRF-SIGMOD-2024}, and face recognition~\cite{AnalyticDB-VLDB-2020}.

\begin{table*}[!t]
\begin{footnotesize}
  \centering
\vgap\caption{We compare our methods with existing solutions.
Among them, $\PRE$, $\POST$, and $\SUPER$ primarily focus on search within a single graph index.
$\SERF$\cite{SeRF-SIGMOD-2024} is a compression-based method, while $\SEG$\cite{Window-Filter-ICML-2024} and $\IRANG$~\cite{iRangeGraph-SIGMOD-2025} are reconstruction-based approaches.
The Scalability and Query Efficiency are confirmed by our experimental study. 
Theoretical Guarantee indicates whether the methods provide query time complexity analysis. 
Adaptability assesses if the methods can extend beyond graph-based indexing (for each range). Additionally, Half-Bounded RFAKNN is to check whether tailored methods are designed specifically for this type of query.}\vgap \vgap
  \begin{tabular}{l|c|ccccccc}
    \toprule
    Feature         & $\HBI$ (Our) & $\PRE$ & $\POST$ & $\SUPER$~\cite{Window-Filter-ICML-2024} &  $\SERF$~\cite{SeRF-SIGMOD-2024}& $\SEG$~\cite{Window-Filter-ICML-2024} & $\IRANG$~\cite{iRangeGraph-SIGMOD-2025} \\ 
    \midrule
    Scalability          & $\star\star\star$  & $\star\star\star$ & $\star\star\star$ & $\star\star$      & $\star\star\star$    & $\star\star\star$ & $\star\star$\\ 
    Query Efficiency    & $\star\star\star$  & $\star$           & $\star$           & $\star\star\star$ & $\star\star$         & $\star\star$      & $\star\star\star$ \\
    Theoretical Guarantee & \checkmark         & $\times$          & $\times$          & $\times$          & $\times$             & \checkmark        & $\times$\\
    Adaptation         & \checkmark         & \checkmark        & \checkmark        & \checkmark        & $\times$             & \checkmark        & $\times$\\ 
    Half-Bounded RFAKNN             & \checkmark         & $\times$          & $\times$          & $\times$          & \checkmark           & $\times$          & $\times$\\ 
    \bottomrule
  \end{tabular}
  \label{tab:features}
\end{footnotesize}\vspace{-0.25em}
\end{table*}

\stitle{Search Principles.}
To process RFAKNN queries, adopting a graph index (such as $\HNSW$) originally designed for AKNN search is a straightforward approach. 
The main challenge lies in handling the additional query ranges and two commonly used principles for this are:
\underline{1) $\PRE$}, which filters out points whose numeric values fall outside the query range. 
Since it ensures that only in-range points are included, the resulting graph index may become less-connected after removing out-of-range points. 
Conducting RFAKNN searches on such a sparse graph often leads to reduced query accuracy.
\underline{2) $\POST$}, which retains all points, including those that fall outside the query range. 
It progressively identifies points near the query point but only adds a point to the result if its numeric values fall within the query range. 
Although $\POST$ avoids the sparsity issue, it can operate on a graph index containing a large number of out-of-range points, which can slow down query processing.
While further optimizations such as $\SUPER$~\cite{Window-Filter-ICML-2024} have been proposed to improve query speed, these often come at the cost of increased space requirements.

\stitle{State-of-the-art.}
The aforementioned methods typically assume the creation of a \textit{single} graph index for all points in $\mathcal{D}$, which is then adapted for RFAKNN searches using either the $\PRE$ or $\POST$ principles.  
To address the limitations of these methods, more advanced techniques have been developed that focus on creating \textit{multiple} graph indexes.
Recall that, in RFAKNN, users can specify up to $O(N^2)$ possible query ranges $[l_q, r_q]$, as both $l_q$ and $r_q$ are bounded by the cardinality $N$ of the database.
If we create the graph index for each possible range, then when a query with range $[l_q, r_q]$ is issued, the precomputed graph index corresponding to range $[l_q, r_q]$ is utilized for query processing, which ensures both query accuracy and efficiency.
Yet, maintaining $O(N^2)$ graph indexes demands an exhaustive amount of storage.

To avoid the need for materializing $O(N^2)$ graph indexes, existing methods aim to reduce the amount of stored information, and can be categorized into two types:
\underline{1) Compression-based methods~\cite{SeRF-SIGMOD-2024}}, which exploit the relationships between graph indexes of different ranges to enable compression, which achieves an average index size of $O(N \log N)$. 
However, these methods employ lossy compression. As a result, if the index information for a specific query range is not encoded in the compressed representation, query accuracy may degrade.
\underline{2) Reconstruction-based methods~\cite{Window-Filter-ICML-2024,iRangeGraph-SIGMOD-2025}}, which select a subset of ranges from the total $O(N^2)$ possible query ranges to construct graph indexes, and these selected ranges are organized as a segment tree.
When an RFAKNN query with a range $[l_q, r_q]$ is received, the range $[l_q, r_q]$ is decomposed into subranges, which are precisely recorded in the segment tree. 
The graph indexes corresponding to these subranges are then utilized to reconstruct the final graph index for the query over the range $[l_q, r_q]$. 
The segment tree plays a crucial role in recovering unrecorded ranges, ensuring the accuracy of the RFAKNN search query. 
However, this accuracy comes with a cost. 
To reconstruct the range for a query, the process will decompose the range into as many as $O(\log N)$ subranges, which negatively impacts query efficiency.

\stitle{Theoretical Findings.}
An intriguing observation is that both compression-based methods and current reconstruction-based approaches adhere to the $\PRE$ principle: they construct the graph index only after excluding \textit{all} out-of-range points. 
However, employing $\PRE$ introduces drawbacks: compression-based methods fail to guarantee query accuracy, and reconstruction-based approaches resolve this limitation but at the expense of reduced query efficiency.
This raises a critical question: is it possible to achieve both query accuracy and efficiency simultaneously?
Surprisingly, our findings reveal that adopting the $\POST$ principle provides a promising solution to this question. 
$\POST$ ensures query accuracy by including all data points during the search. 
Yet, this accuracy comes at the cost of increased query overhead due to the inclusion of \textit{all} out-of-range points in $\mathcal{D}$, making a direct application of $\POST$ \textit{inefficient}.
Recall that $\POST$ operates over the entire range $[1, N]$. 
To address this, we investigate the behavior of the graph index when applied to any superset range $[a, b]$ of the query range $[l_q, r_q]$, where $l_q \ge a$ and $r_q \le b$, rather than only on range $[1, N]$ used in $\POST$.
In this case, the graph index for the superset $[a, b]$ includes all in-range points as well as \textit{some} out-of-range points for query range $[l_q, r_q]$. 

Our first theoretical finding is that, for a query range $[l_q, r_q]$, using a graph index corresponding to a superset range $[a, b]$ can correctly answer the RFAKNN query for $[l_q, r_q]$.
This also explains why $\POST$ guarantees query accuracy (but over-killed): it builds the index over the entire range $[1, N]$, which is the superset of all possible query ranges.
Our second theoretical finding is that, if the size of the superset range $[a, b]$ is only $\beta$ times larger than the query range $[l_q, r_q]$, the query time complexity increases by merely a linear factor compared to the original search time. 
This insight paves the way for a more efficient alternative: Instead of using the graph index for the full range $[1, N]$ as required by $\POST$, we propose using a graph index corresponding to a tight superset range $[a, b]$ that is only slightly larger than the range $[l_q, r_q]$.
In this way, we preserve the accuracy guarantees of $\POST$ while improving query efficiency.

\stitle{Our Novel Methods.}
When combining the aforementioned theoretical findings with current reconstruction-based methods, we conceived the idea for our novel design: Elastic Graph ($\kw{ESG}$), a new method for RFAKNN search.
We first address the problem of answering half-bounded RFAKNN queries, where the input range has the form $[1, r_q]$, with $r_q \in [1, N]$ being the only flexible variable.
We show that, by creating graph indexes for only $\log N$ ranges of the form $[1, N/2^i]$, where $i \in [0, \log N]$, RFAKNN queries with range $[l_q, r_q]$ can be efficiently resolved by referencing its tightest superset.
We also prove that the selected superset range is at most twice the size of the query range, ensuring query efficiency.
Also, instead of constructing the graph indexes for these $\log N$ ranges $[1, N/2^i]$ individually, we propose a holistic indexing approach that constructs a single range that generates the index for all necessary subranges, streamlining the indexing process.

Next, we extend our method to support general RFAKNN queries, where the query range $[l_q, r_q]$ includes two flexible variables, $l_q$ and $r_q$.  
Our approach also leverages a segment tree, with a key innovation in query processing: instead of identifying subranges that \textit{exactly} match the query range, we allow the combined results of subranges to form a superset of the input range.  
We theoretically demonstrate that this elastic partitioning ensures the query range can be divided into at most \textit{two} subranges, thereby eliminating the $O(\log N)$ factor present in the query processing of existing reconstruction-based solutions.  
Furthermore, we propose a novel method for constructing graph indexes for all ranges in the segment tree, significantly accelerating index construction.
Table~1 presents a comparison between our proposed method and the existing approaches for RFAKNN queries.

\stitle{Contributions.}
We summarize our contributions as follows:

\vspace{0.5em}
\sstitle{Problem Analysis of State-of-the-art (\S~\ref{sec:Preliminary}).}
We classify existing methods for answering RFAKNN queries into two main categories. 
We find that compression-based methods often compromise on query accuracy, whereas reconstruction-based methods suffer from query inefficiencies. 
This motivates our investigation into novel approaches for processing RFAKNN queries.

\vspace{0.5em}
\sstitle{Theoretical Findings (\S~\ref{sec:Theoretical}.)}
To balance query accuracy and efficiency, we leverage the $\POST$ principle and investigate the relationship between a graph index constructed on a superset of a range (to accommodate out-of-range points) and the original range. 
Our findings demonstrate that a graph index built on the superset range ensures accurate query processing when the original range is queried. 
Moreover, we establish that the query time for the graph constructed on the superset range is bounded by a constant factor of the query time for the original range.
These results extend the existing $\POST$ principle, laying the groundwork for the development of our novel methods.

\vspace{0.5em}
\sstitle{Novel RFAKNN Query Processing Methods (\S~\ref{sec:Elastic}).}
We first propose a novel index structure tailored for the half-bounded RFAKNN search, based on our theoretical findings.
This index is constructed over $\log N$ ranges and is proven to be sufficient for answering range queries for all input ranges.
We then design a new index structure for the general RFAKNN search, leveraging a segment tree approach.
Our main technical contribution lies in introducing a novel query-processing method.
This method ensures that the query range is divided into at most two subranges in the worst case.

\vspace{0.5em}
\sstitle{Extensive Experimental Studies (\S~\ref{sec:Experiments}).}
We evaluated our proposed methods on several real-world datasets and compared them with state-of-the-art methods. 
The experimental results demonstrate that our approaches outperform existing methods in both query efficiency and accuracy. 
Moreover, our methods are scalable to datasets with up to 100 million data points, making it desirable for large-scale RFAKNN searches.

Due to space limitations, some proofs and experiments can be found in our technical report~\cite{technicalreport}. 

\begin{table}[t]
  \caption{A Summary of Notations}\vgap\vgap
  \label{tab:notation}
  \small
  \begin{tabular*}{\linewidth}{@{\extracolsep{\fill}} p{15mm} | p{70mm}}
    \toprule
    Notation   &  Description\\
    \midrule
    $\mathcal{D}$ &  A set of vectors/points with numerical attributes \\
    $N$        &  The cardinality of $\mathcal{D}$ \\
    $q$        &  The query vector \\
    $e$        &  The elastic factor \\
    $d$        &  The dimensionality of $\mathcal{D}$ \\
    $\mathcal{N}(u)$ & The neighbors of node $u$ in a graph index \\
    $[l,r]$    &  The range with left bound $l$ and right bound $r$ \\
    $|[l,r]|$    &  The size of range $[l,r]$ \\
    $R$        &  The set of ranges\\
    $\mathbb{R}^d$  &  The $d$-dimensional Euclidean space\\
    $\|u,v\|$  &  The Euclidean distance between $u$ and $v$ \\
    $\ell$    &  The range length bound \\
    $\mathcal{I}$  &  The graph index such as $\HNSW$  \\
    $\HBIO, \HBIT$  &  The elastic graph indexes\\
    \bottomrule
  \end{tabular*}
\end{table}

\section{Preliminary}\label{sec:Preliminary}
We begin by introducing the problem of the approximate $k$-nearest neighbors (AKNN) search and its graph-based solutions. 
Following this, we formally define the range-filtering AKNN search problem and provide an overview of existing approaches. 
For clarity, the commonly used notations are summarized in Table~\ref{tab:notation}.

\subsection{AKNN Search and Graph-Based Solutions}
The $k$-nearest neighbors (KNN) search in high-dimensional spaces involves retrieving the top $k$ closest vectors to a given query point based on a specified distance metric.  
This study takes the Euclidean distance as an instance, as the methods used for handling Euclidean distance can be extended to other distance metrics.

\begin{definition}\label{def:anns}
Let $\mathcal{D} = \{v_1, v_2, \ldots, v_N\}$ be a set of $N$ points/vectors in $\mathbb{R}^d$.
Given a query point $q \in \mathbb{R}^d$ and an integer $k > 0$, the \textbf{k-nearest neighbors (KNN) query} returns the set of $k$ points $S \subseteq \mathcal{D}$ with the smallest Euclidean distances to $q$. 
Formally, $\forall v \in \mathcal{D} \setminus S$, $||q - u|| \le ||q - v||$ for all $u \in S$.
\end{definition}

Due to the curse of dimensionality~\cite{Curse-of-dim-1998}, searching for the exact KNN of a query point in high-dimensional space is computationally expensive. 
Therefore, a relaxed version of the KNN search, known as the \textbf{Approximate K-Nearest Neighbors (AKNN)} search, has been proposed to return the Approximate KNN for a query point.

\stitle{Graph Index.}
To process AKNN queries, numerous methods have been proposed. 
Among these, graph-based algorithms~\cite{FANNG:harwood2016fanng,DBLP:journals/pvldb/NSGFuXWC19,DBLP:journals/pami/SSGFuWC22,tMRNG:journals/pacmmod/PengCCYX23,HVS:DBLP:journals/pvldb/LuKXI21,Diskann-NIPS-2019} have gained great attention due to their state-of-the-art search performance. 
These algorithms construct a graph $G$ as an index, where all points in $\mathcal{D}$ are represented as vertices, and edges are carefully selected to satisfy the monotonic search network (MSNET) property~\cite{DBLP:journals/pvldb/NSGFuXWC19}. 
Specifically, the graph index leverages an edge occlusion strategy~\cite{FANNG:harwood2016fanng,DBLP:journals/pami/HNSWMalkovY20,DBLP:journals/pvldb/NSGFuXWC19,tMRNG:journals/pacmmod/PengCCYX23,Diskann-NIPS-2019,SA-tree-VLDBJ-2002-navarro} to prune redundant edges while maintaining the essential MSNET properties, which underpin the effectiveness of these methods. 
Additionally, theoretical studies demonstrate that the degree of graph $G$ after edge occlusion remains bounded by a constant, ensuring both efficiency and scalability.

\begin{algorithm}[!t]
	\caption{Graph-Search$(G,ep,q,m)$}
	\label{alg:graph-search}
	\begin{small}
        $P \gets ep$ \tcp*{priority queue, min heap by distance}
        $Q \gets ep$ \tcp*{result queue, max heap by distance}
        \While{$P \neq \emptyset$}{
            $u \gets P.top()$\;
            \textbf{if} $||q, u|| > ||q, Q.top()||$ \textbf{then} break\;
            
            \For{$v \in N(u)$}{
                \textbf{if} $v$ is visited \textbf{then} continue\;

                //$\PRE$: \textbf{If} $v$ is out-of-range \textbf{then} Continue\;
                
                $P \gets P \cup v$\;

                //$\POST$: \textbf{If} $v$ is out-of-range \textbf{then} Continue\;
                
                $Q \gets Q \cup v$\;
            }
            \If{$Q.size()>m$}{
               $Q.resize(m)$\tcp{keep top $m$ results}
           }
        }
        \Return{$Q$}
        \end{small}
\end{algorithm}

\stitle{Search Algorithm.}
When the graph index $G$ (e.g., $\HNSW$~\cite{DBLP:journals/pami/HNSWMalkovY20}) is created, Algorithm~\ref{alg:graph-search} shows the process of searching for AKNN of a query point $q$.
Initially, the entry point $ep$ is inserted into two heaps, $P$ and $Q$, where $P$ is used to set priorities and $Q$ is used to store the current top-$m$ closest neighbors (Line~1–2), where $m \ge k$ is the beam search size.
Algorithm~\ref{alg:graph-search} proceeds by repeatedly popping a point $u$ from $P$ until $P$ becomes empty (Line~3–4).
For each point $u$, its distance to the query point $q$ is compared with the largest distance between $q$ and the points in $Q$. 
If $u$'s distance is greater than this largest distance, the search stops (Line~5).
Otherwise, for each neighbor $v \in \mathcal{N}(u)$ of $u$ in $G$ (Line~6), it checks whether $v$ has already been visited (Line~7). 
If visited, it is skipped; otherwise $v$ is inserted into both $P$ and $Q$ (Line~9, 11).
When the size of the answer queue $Q$ exceeds $m$, the algorithm adjusts by removing points with larger distances to $q$ to maintain the desired size (Line~12–13).
The top-$k$ closest points in $Q$ are then returned as the answer (Line~14).

\subsection{RFAKNN Search and Existing Solutions}
When a point has an associated numeric attribute (e.g., price), the problem of the AKNN search can be more flexible. 
This flexibility arises because we can pre-assign a range filter in the query to eliminate points whose numeric values do not fall within the specified range. 
In light of this, we define the \textbf{Range-Filtering AKNN (RFAKNN)} search problem as follows.

\begin{definition}\label{def:rfanns}
Let $\mathcal{D} = \{v_1, v_2, \ldots, v_N\}$ be a set of $N$ points in $\mathbb{R}^d$, where each point $v_i$ is associated with a numerical value $i \in [1, N]$.
The \textbf{RFAKNN query} is defined as $Q = (q, l, r)$, where $q$ is a vector and $[l, r]$ denotes a range.
The RFAKNN query returns the result of AKNN search of $q$ in $\mathcal{D}_{[l,r]}$, where $\forall v_i \in \mathcal{D}_{[l,r]}$, $i \in [l, r]$.
\end{definition}

\begin{figure}[!htb]
    \centering
    \includegraphics[scale=0.8]{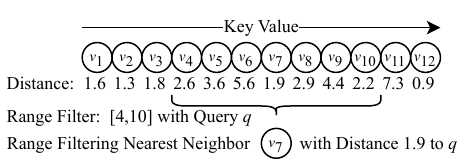}
    \vgap\vgap\caption{The example of the RFAKNN query, where the points $v_i$ has an additional numerical attribute $i$. 
    Given a query point $q$ with the range $[4, 10]$, the answer (for $k=1$) is $v_7$ because the distance from $q$ to $v_7$ is the smallest among all in-range points $\mathcal{D}_{[4,10]}=\{v_4, v_5, \cdots, v_{10}\}$.}
    \label{fig:RFKNNS}
\end{figure}

\stitle{$\PRE$ and $\POST$ Principles.}
There are two principles for answering RFAKNN search queries. 
Both methods build upon the search algorithm presented in Algorithm~\ref{alg:graph-search}, which was originally designed for AKNN queries.

\noindent
$\bullet$ $\PRE$: Using this principle, we identify and discard out-of-range points during the search (Line~9 of Algorithm~\ref{alg:graph-search}). 
Specifically, if the numeric value of a point $v$ is not within the query range $[l, r]$, then $v$ is excluded from the search process (i.e., it is not added to the queue $P$). 
In this way, the RFAKNN query under range $[l, r]$ can be answered by ensuring that only in-range points are considered.

\noindent
$\bullet$
$\POST$: This principle does not discard out-of-range points during the search phase. 
Instead, only when a point is encountered and evaluated for inclusion in the result queue $Q$, it discards the point if its numeric value is not within the query range $[l, r]$ (Line~10 of Algorithm~\ref{alg:graph-search}).

\begin{figure}[!t]
    \centering
    \includegraphics[scale=0.8]{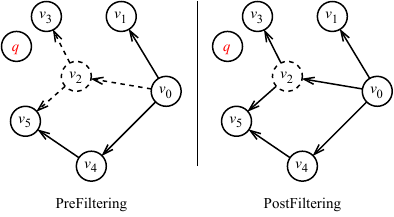}
    \vgap\caption{The Illustration of $\PRE$ and $\POST$}\vgap\vgap
    \label{fig:graph-deleted}
\end{figure}

\begin{exmp}
In Fig.~\ref{fig:graph-deleted}, let $v_0$ be the entry node and $q$ be the query point. 
Assume $v_2$ is an \textit{out-of-range} point for a given RFAKNN query. 
According to the $\PRE$ principle, all edges connected to $v_2$ (depicted as dashed lines) are removed from the graph index $G$.
This removal results in the disconnection between $v_0$ and the nearest neighbor of $q$, $v_3$. 
Thus, the search algorithm returns $v_5$, traversing the path $v_0 \rightarrow v_4 \rightarrow v_5$.
In contrast, the $\POST$ principle retains the edges connected to $v_2$ for searching purposes, while excluding $v_2$ itself as a candidate result.
Thus, it successfully identifies the nearest neighbor, $v_3$, by traversing the path $v_0 \rightarrow v_2 \rightarrow v_3$, leveraging the outgoing edges of $v_2$.
\end{exmp}

\sstitle{Drawbacks.}
The $\PRE$ principle is effective in removing out-of-range points from the graph, which reduces the search space. 
As a result, $\PRE$ ensures high query efficiency. 
However, since $\PRE$ removes some necessary points from the graph, it may cause the graph to become sparse, potentially preventing the search from finding the correct result.
In contrast, $\POST$ directly searches on the whole graph, maintaining its high accuracy. 
Nevertheless, it requires computing the distances of many out-of-range points, which can result in a longer query time.

\stitle{State-of-the-art.}
Recall that $\PRE$ and $\POST$ only work on a \textit{single} graph index created for the entire range $[1, N]$ (i.e., all points in the whole database $\mathcal{D}$).
To overcome their limitations, existing solutions primarily focus on creating \textit{multiple} graph indexes, and these solutions can be categorized into two types.

\noindent
$\bullet$
Compression-based methods. 
These methods conceptually construct the graph index for each of the $O(N^2)$ possible ranges.
We denote that a graph index is created for a range $[l,r]$ when the graph is actually created for points in $\mathcal{D}_{[l,r]}$, where $l, r \in [1, N]$. 
When an RFAKNN query is issued, the graph corresponding to the query range $[l_q,r_q]$ is retrieved, and the query is processed using a graph search algorithm (e.g., Algorithm~\ref{alg:graph-search}).
Since materializing and storing all $O(N^2)$ possible ranges is infeasible, these methods propose a holistic index that compactly compresses these $O(N^2)$ indexes to a size of $O(N \log N)$.
However, current compression-based methods are lossy, meaning that the information corresponding to a range $[l,r]$ may be incomplete. 
As a result, queries for such ranges may lack accuracy.
For instance, the RFAKNN query performance deteriorates if the length of a query range is small, making it difficult to achieve a recall of 0.8 or higher.

\begin{figure}[!t]
    \centering
    \includegraphics[scale=0.6]{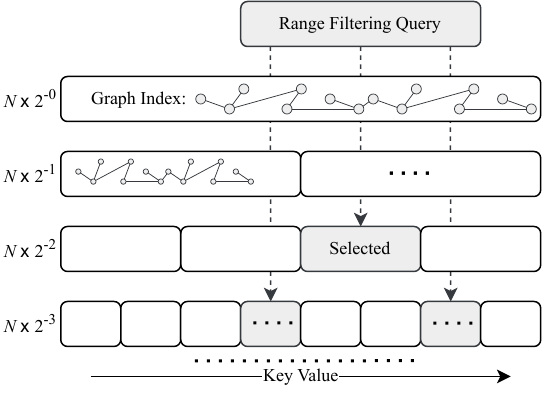}
    \vgap\caption{The Example of Reconstruction-based Methods for the RFAKNN Query}\vgap\vgap
    \label{fig:segmengt-tree}
\end{figure}

\noindent
$\bullet$
Reconstruction-based methods.
To improve query accuracy while maintaining reasonable space consumption, reconstruction-based solutions are proposed.
The main idea is that, instead of materializing the index for all possible ranges, a subset of ranges is selected to index. 
Specifically, a segment tree is employed for this selection purpose (see Fig.~\ref{fig:segmengt-tree}). 
Initially, the entire range $[1, N]$ is selected as the root and indexed. 
Subsequently, the range is divided into subranges (for simplicity, we use fanout $2$ as an example in this paper). 
This division continues until the number of points within a given subrange falls below a predefined threshold. 
In total, the segment tree indexes $O(N \log N)$ ranges, which is a subset of the $O(N^2)$ possible ranges.
Unlike compression-based solutions, when a query with a range $[l, r]$ is issued, the segment tree can reconstruct the index even if the query range is not explicitly recorded in the tree. 

As shown in Fig.~\ref{fig:segmengt-tree}, when a query range $[l, r]$ is received, it is divided into multiple subranges that are already recorded in the tree. 
Thanks to the segment tree, this process of dividing the query range into recorded subranges can be performed efficiently. 
Subsequently, the graph indexes for these recorded subranges are combined to generate the index for the query range, thereby improving accuracy through reconstruction. 
However, this process incurs a cost: the indexes for $O(\log N)$ recorded subranges must be combined to answer the RFAKNN query, which can lead to slower query speeds.

\sstitle{Drawbacks.}
Although designed for RFAKNN queries, current state-of-the-art methods~\cite{iRangeGraph-SIGMOD-2025,Window-Filter-ICML-2024,SeRF-SIGMOD-2024} face several challenges (see also Table~1). 
For compression-based methods, while they enable rapid query responses, their lossy nature inherently compromises query accuracy. 
For reconstruction-based methods, although they dynamically refine the range to mitigate accuracy loss, they require dividing the query into $O(\log N)$ subranges, which slows down the query process.
These drawbacks motivate the study of our work.

\section{Theoretical Findings}\label{sec:Theoretical}
This section provides the theoretical findings that underpin our new methods for RFAKNN queries.

\subsection{From $\PRE$ to $\POST$}
As introduced in the previous section, state-of-the-art methods fall short in balancing query accuracy and query speed. 
Interestingly, we observe that both compression-based and reconstruction-based methods rely on the $\PRE$ principle: 
They work only on graph indexes created on the in-range points of the query range, while completely ignoring the out-of-range points.
In light of this, we resort to the $\POST$ principle to explore its potential in addressing the drawbacks of existing methods.
However, this transition is non-trivial as $\POST$ typically requires searching across all points in the database (corresponding to the full range $[1, N]$) for any given query range. 
This results in query inefficiency, as it often includes many points outside the query range.
For this purpose, we provide a theoretical analysis of how the $\POST$ principle can be \textit{enhanced} to ensure query efficiency without compromising query accuracy. 

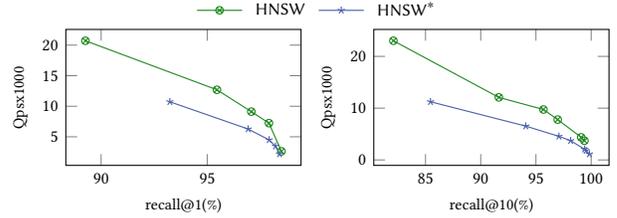
\begin{figure}[t!]
\centering
\begin{small}
\begin{tikzpicture}
    \begin{customlegend}[legend columns=2,
        legend entries={$\HNSW$,$\HNSW^*$},
        legend style={at={(0.45,1.15)},anchor=north,draw=none,font=\scriptsize,column sep=0.1cm}]
    \addlegendimage{line width=0.15mm,color=forestgreen,mark=otimes,mark size=0.5mm}
    \addlegendimage{line width=0.15mm,color=navy,mark=star,mark size=0.5mm}
    \end{customlegend}
\end{tikzpicture}
\\[-\lineskip]

\begin{tikzpicture}[scale=1]
\begin{axis}[
    height=\columnwidth/2.5,
width=\columnwidth/1.8,
xlabel=recall@1(\%),
ylabel=Qpsx1000,
label style={font=\scriptsize},
tick label style={font=\scriptsize},
grid style=dashed,
]
\addplot[line width=0.15mm,color=forestgreen,mark=otimes,mark size=0.5mm]
plot coordinates {
(89.26, 20.7158)
(95.46, 12.692)
(97.08, 9.118)
(97.91, 7.232)
(98.5,  2.622)
};
\addplot[line width=0.15mm,color=navy,mark=star,mark size=0.5mm]
plot coordinates {
(93.24, 10.727)
(96.93, 6.26822)
(97.92, 4.48905)
(98.22, 3.46345)
(98.41, 2.17859)
};

\end{axis}
\end{tikzpicture}\hspace{1mm}
\begin{tikzpicture}[scale=1]
\begin{axis}[
    height=\columnwidth/2.5,
width=\columnwidth/1.8,
xlabel=recall@10(\%),
ylabel=Qpsx1000,
label style={font=\scriptsize},
tick label style={font=\scriptsize},
grid style=dashed,
]
\addplot[line width=0.15mm,color=forestgreen,mark=otimes,mark size=0.5mm]
plot coordinates {
(82.09, 23.018)
(91.64, 12.080)
(95.68, 9.740)
(96.97, 7.791)
(99.1, 4.364)
(99.39, 3.711)
};
\addplot[line width=0.15mm,color=navy,mark=star,mark size=0.5mm]
plot coordinates {
(85.45, 11.235)
(94.08, 6.533)
(97.08, 4.556)
(98.14, 3.716)
(99.41, 2.058)
(99.54, 1.800)
(99.84, 1.138)
};

\end{axis}
\end{tikzpicture}\hspace{2mm}
\vgap\vgap\caption{The Illustration of the Potential of $\POST$}\vgap\label{fig:Qps-filter-half}
\end{small}
\end{figure}

\begin{exmp}
Figure~\ref{fig:Qps-filter-half} shows the effectiveness of utilizing $\POST$ when half of the data points are out-of-range.
For this evaluation, we use $\HNSW$ to construct the graph index and assess its performance on the SIFT dataset. 
For processing RFAKNN queries, the $\HNSW$ method, represented by the green line, builds the index exclusively from in-range points. 
Conversely, the $\HNSW^*$ method constructs an $\HNSW$ index for all points and integrates the $\POST$ method during search operations.
The results reveal that $\HNSW^*$ with $\POST$ achieves performance levels comparable to $\HNSW$, particularly at high recall levels. 
Even at moderately lower recall levels, the performance difference remains minimal, within a twofold range.
\end{exmp}

\stitle{Elastic Factor.}
For $\POST$, for any query range $[l, r]$, it operates on the graph containing all points (i.e., using only \textit{one} range $[1, N]$) in $\mathcal{D}$.
This is the root cause of $\POST$ being slow: it must process a large number of out-of-range points. 
To address this limitation, we propose relaxing this requirement by assuming that graph indexes are created for multiple ranges, denoted as $R$. 
We will analyze the performance of $\POST$ when multiple ranges are applied. 
The relationship between a range $[l, r]$ and the set of ranges $R$ is captured using the concept of Elastic Factor.

\begin{defn}[Elastic Factor]\label{def:elastic_factor}
        Given a set of ranges $R$, where each range $[l_i, r_i] \in R$ has $l_i, r_i \in [1, N]$, the elastic factor of a range $[l, r]$, with $l, r \in [1, N]$, is defined as:
\begin{equation*}
    e(R,[l,r]) = \max_{[l,r] \subset [l_i,r_i]}\left(\frac{|[l,r]|}{|[l_i,r_i]|} \right).
\end{equation*}
Here, $|[l, r]| = r - l + 1$ represents the length of a range $[l, r]$.
\end{defn}

The elastic factor $e(R, [l, r])$ can be understood as selecting the closest superset of a query range $[l, r]$ from $R$ to measure the ratio. 
This concept is similar to the blowup factor~\cite{Window-Filter-ICML-2024}, but it focuses on how the range set covers the query range. 
Moreover, current studies on $\POST$ only consider when $R$ include one singe range $[1, N]$, which results in a very small elastic factor $e(R, [l, r])$.

\subsection{Findings}
Based on the elastic factor, we provide two key findings.
\textbf{Our first theoretical finding} is that when $R$ includes a superset range $[a,b]$ of the query range $[l, r]$, the query accuracy is not compromised.
This observation is intuitive: when $[a, b] \in R$ is a superset of $[l, r]$, all in-range points for the query range $[l, r]$ are also included in the range $[a, b]$.
This implies that the graph index created for the query range $[l, r]$ is a subgraph of the superset range $[a, b]$. 
According to the properties of the monotonic search network, KNN searches on the subgraph can also be performed within the superset graph. Thus, we confirm this finding.
This observation also explains why creating a graph index for range $[1, N]$ alone is sufficient for $\POST$, as $[1, N]$ serves as a superset for all possible query ranges.

\vspace{0.5em}
\textbf{Our second theoretical finding} is that when $e(R, [l, r])$ is sufficiently large (as $R$ can now include more ranges), the query speed of $\POST$ can be improved without sacrificing accuracy.
Below is the formal proof of this claim.

\stitle{Complexity When $k$ is 1.}
To investigate the query complexity of $\POST$ under elastic factor constraints, we begin by examining the query complexity of existing graph search methods (e.g., Algorithm~\ref{alg:graph-search}) without range filtering, with a particular focus on the time complexity of $k$-nearest neighbor (KNN) search.
In the existing literature, the special case of $k = 1$ has been extensively studied.
Specifically, these methods construct graph indexes that satisfy the Monotonic Search Network (MSNET) property using edge-pruning strategies, ensuring that each point in the graph is associated with a monotonic search path.

\vspace{-0.5em}
\begin{defn}[Monotonic Search Path~\cite{DBLP:journals/pvldb/NSGFuXWC19}]
\label{def:monotonic_search_path}
    Let $\mathcal{D}$ be a database of $N$ points in $\mathbb{R}^d$, and let $G$ be a graph defined on $\mathcal{D}$. 
    For any two points $p, q \in \mathcal{D}$, let $v_1, v_2, \ldots, v_k$ denote a path from $p$ to $q$ in $G$. This path is called a monotonic path if and only if $\forall i \in [1, k-1]$, $||v_i - q|| > ||v_{i+1} - q||$.
\end{defn}

Building on the concept of a monotonic search path, a Monotonic Search Network is defined as follows:

\vspace{-0.5em}
\begin{defn}[Monotonic Search Network~\cite{DBLP:journals/pvldb/NSGFuXWC19}]
\label{def:msnet}
    Given a database $\mathcal{D}$ of $N$ points in $\mathbb{R}^d$ and a graph $G$ defined on $\mathcal{D}$, the graph $G$ is called a Monotonic Search Network if and only if there exists a monotonic search path between every pair of points $p, q \in \mathcal{D}$.
\end{defn}

\vspace{-0.5em}

From the above definitions, the query complexity of KNN search when $k$ is $1$ is derived~\cite{DBLP:journals/pvldb/NSGFuXWC19}:

\vspace{-0.5em}
\begin{equation}
\label{eq:nn_time_complexity}
    O\left(\frac{N^{\frac{1}{d}} \log N^{\frac{1}{d}}}{\bigtriangleup r}\right),
\end{equation}

Here, $\bigtriangleup r$ represents the minimal distance required to move closer to the query, while the degree of the graph index is bounded by a constant~\cite{DBLP:journals/pvldb/NSGFuXWC19}. 
This query complexity can be interpreted as the expected length of the monotonic search path within the graph.

\stitle{Complexity for General $k$.}
Equation~\ref{eq:nn_time_complexity} only accounts for the complexity where $k = 1$, leaving the query complexity for other cases unexplored. 
To bridge this gap, we offer the following conclusion:

\begin{thm}\label{thm:KNN-time-complexity}
    Under the same assumptions as in~\cite{DBLP:journals/pvldb/NSGFuXWC19}, the expected search path length of MSNET with reverse edges for KNN search is:
    \begin{equation}\label{eq:top-k-time-complexity}
        O\left(\frac{N^{\frac{1}{d}} \log N^{\frac{1}{d}}}{\Delta r} + k\right).
    \end{equation}
\end{thm}

\stitle{Linking with Range Constraints.}
Next, we focus on the query complexity of $\POST$ for range-filtering $k$-nearest neighbor (RFKNN) search, as this paper addresses scenarios where queries involve a range filter.
We show that the RFKNN search problem can be identically solved by performing an $h$-nearest neighbor search, where $h$ is a value no smaller than $k$.

\begin{lem}\label{lem:top-k}
Given a database $\mathcal{D}$ of $N$ points in $\mathbb{R}^d$ with an additional numerical attribute, a query $q$ with a filter $[l, r]$, and the $h$-nearest neighbor set $S_h$ of $q$, let $S_k$ represent the range-filtering $k$-nearest neighbor set of $q$. 
If $|\{v_i \in S_h \mid i \in [l, r]\}| \geq k$, then $S_k \subseteq S_h$.
\end{lem}

Lemma~\ref{lem:top-k} shows that the range-filtering $k$-nearest neighbor set, denoted as $S_k$, can be fully contained within the $h$-nearest neighbor set $S_h$ of $q$, provided that the number of in-range points in $S_h$ is no smaller than $k$.
Building on this observation, an existing graph index can be utilized to perform an incremental $h$-nearest neighbor search. 
The process continues until $S_h$ contains at least $k$ in-range points, at which point $S_h$ is returned as the result for the RFKNN search. 
If this condition is not met, the search progresses incrementally to include the $(h+1)$-neighbor, which is how $\POST$ works.

\stitle{Complexity Under Range Constraints.}
Since the query complexity for KNN search is already provided in Equation~\ref{eq:top-k-time-complexity}, the next challenge is estimating the value of $h$, which determines the query complexity for RFKNN search (by substituting $k$ with $h$ in the equation).
In the worst case, $h$ could equal $N$, as it may be necessary to process all data points to report the result for range-filtering $k$-nearest neighbor search. 
However, by incorporating the elastic factor, we can prove that the expected value of $h$ is bounded by $\frac{k}{c}$, where $c$ is a value no smaller than the elastic factor of the query range $[l, r]$. 
This result ensures that the complexity of the RFKNN search remains comparable to the original KNN search, with only a proportional factor introduced by the range-filtering mechanism.

\begin{thm}\label{thm:half-blood-search-time-complexity}
    Given a database $\mathcal{D}$ of $N$ points, a range set $R$ (where each range is built using an MSNET), and a query $q$ with a filter range $[l, r]$. Assuming the conditions outlined in~\cite{DBLP:journals/pvldb/NSGFuXWC19} and that the attribute values of points in $\mathcal{D}$ are independent, if the elastic factor of $[l, r]$ satisfies $e(R, [l, r]) \geq c$ for some constant $c$, the MSNET with reverse edges returns the range-filtering $k$ nearest neighbors with an expected search path length:
    
\begin{equation}\label{eq:half-blood-time-complexity}
    O(\frac{N^{\prime\frac{1}{d}} \log N^{\prime\frac{1}{d}}}{\bigtriangleup r^{\prime}} + k/c)
\end{equation}
     where $N^{\prime}=|[l,r]|/c$ and $\bigtriangleup r^{\prime}$ is the minimal distance to get closer to the query.
\end{thm}

Theorem~\ref{thm:half-blood-search-time-complexity} shows that $\POST$ for RFKNN queries exhibits time complexity comparable to the optimal approach when the elastic factor is sufficiently large. 
Specifically, the time complexity described in Equation~\ref{eq:half-blood-time-complexity} can be reformulated as $O(N^{\prime\frac{2}{d}}\log N^{\prime})$ with a probability of at least $1-(1/e)^{\frac{d}{4}(1-\frac{3}{e^2})}$ with the conclusion in~\cite{tMRNG:journals/pacmmod/PengCCYX23}. 
The size of the index range, $N^{\prime}$, is at most $1/c$ of $N^{\prime\prime}$, which implies that the time complexity increases only by a constant factor. Consequently, it can be further reduced to $O(N^{\prime\prime\frac{2}{d}}\log N^{\prime\prime})$. 
The next challenge lies in designing a method to generate the range set $R$ such that the elastic factor exceeds a constant threshold $c$ (e.g., 0.5) for any query range.

\section{Our RFAKNN Query Processing Methods}\label{sec:Elastic}
In the previous section, we provided a theoretical analysis to ensure that, under an elastic factor no smaller than $c$, the search method based on the $\POST$ principle remains efficient.
In this section, we propose novel reconstruction-based methods that fully leverage this conclusion to process RFAKNN queries efficiently. 
We first design a solution for the half-bounded RFAKNN search, and then we discuss the solution for the general RFAKNN search case.
Note that our initial focus is on the case where $c = \frac{1}{2}$ and the more general case is discussed as an extension.

\subsection{The Method for Half-Bounded Queries}\label{subsec:HBI-1D}
To make it easy to understand, we first consider half-bounded RFAKNN queries. 
A half-bounded query means that the query range takes the form of either $[1, r]$ or $[r, N]$, where $r \in [1, N]$. 
This occurs when either the left bound is fixed at $1$ or the right bound is fixed at $N$. 
For simplicity, we discuss the case of $[1, r]$, as the case of $[r, N]$ is similar.
To answer the half-bounded RFAKNN query, an intuitive approach is to create $N$ graphs, one for each range $[1, i]$, where $i \in [1, N]$. 
However, due to the elastic factor, it is sufficient to create only $O(\log N)$ graphs. 
This results in the $\HBIO$ index, which we define as follows.

\begin{definition}
Given a database $\mathcal{D}$ of $N$ points, the $\HBIO$ index contains $\log N$ graphs. Each graph is constructed for the range $[1, N/2^i]$, where $i \in [0, \log N]$. 
\end{definition}

\begin{figure}[!t]
    \centering
    \includegraphics[scale=0.6]{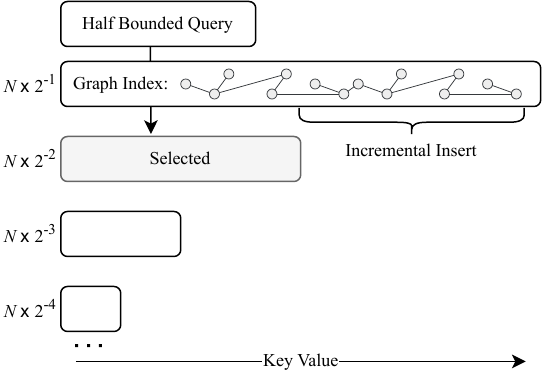}\vgap
    \vgap\caption{The Example of $\HBIO$}\vgap
    \label{fig:HBI-1D-example}
\end{figure}

\begin{example}
Fig.~\ref{fig:HBI-1D-example} shows $\HBIO$, which consists of $O(\log N)$ graphs, each corresponding to range $[1, N/2^i]$, where $i \in [0, \log N]$.
\end{example}

\stitle{Query Processing.}
To process an RFAKNN query with point $q$ and range $[1, r]$, we retrieve the $\HBIO$ index to locate the recorded graph under the range $[1, N/2^{\lceil \log r \rceil}]$. 
Then, we utilize $\POST$ to search this graph to find the answer.
The rationale and justification for using this range, $[1, N/2^{\lceil \log r \rceil}]$, are as follows:

\begin{lemma}\label{lemma:half}
    $[1, r] \subseteq [1, N/2^{\lceil \log r \rceil}]$, and $\frac{|[1, r]|}{|[1, N/2^{\lceil \log r \rceil}]|} \geq 0.5$. 
\end{lemma}

In Lemma~\ref{lemma:half}, the condition that $[1, r] \subseteq [1, N/2^{\lceil \log r \rceil}]$ ensures that the range $[1, N/2^{\lceil \log r \rceil}]$ is a superset of $[1, r]$, which guarantees the query accuracy.
The condition $\frac{|[1, r]|}{|[1, N/2^{\lceil \log r \rceil}]|} \geq 0.5$ ensures that the elastic factor of range $[1, r]$ is at least $0.5$, which supports efficient query response.
From Lemma~\ref{lemma:half}, we conclude that $[1, N/2^{\lceil \log r \rceil}]$ serves as an appropriate superset of $[1, r]$, balancing accuracy and efficiency for RFAKNN queries.

\begin{algorithm}[!t]
	\caption{Build-1D-Index$(\mathcal{D},N)$}
	\label{alg:build-1D-index}
	\begin{small}
	\KwIn{point $u\in \mathcal{D}$, cardinality $N$}
	\KwOut{index $\HBIO$}
        $R \gets [1, N/2^i]$ where $i \in [0, \log N]$\tcp*{init ranges}
        $\HBIO \gets \emptyset,\mathcal{I} \gets \emptyset$\tcp*{init index}
        $pre \gets 1$\;
        \For{$cur \gets 1$ \textbf{to} $N$}{
            \If{$[1,cur] \in R$}
            {   
                insert $v_i$ into the graph $\mathcal{I}$, $i\in [pre, cur]$\;
                $\HBIO \gets \HBIO \cup (\mathcal{I}, [1,cur])$\;
                $pre \gets cur$\;
            }
            
        }
        \Return{$\HBIO$}
        \end{small}
\end{algorithm}

\stitle{Index Construction.}
Recall that it is feasible to construct the graph index separately for all ranges $R = \{[1, N/2^i] \mid i \in [0, \log N]\}$. 
However, we observe that when constructing the graph index for the range $[1, N]$, all other subranges in $R$ can be derived as byproducts.
Specifically, by sorting the points in non-increasing order based on their attribute values, the index for any range $[1, \textit{cur}]$ (where $\textit{cur} \leq N$) is incrementally created until we reach $\textit{cur} = N$ and the index for $[1, N]$ is created.

\sstitle{Algorithm.}
Algorithm~\ref{alg:build-1D-index} outlines the process for creating $\HBIO$. 
Initially, we determine the required ranges $R$ for the $\HBIO$ (Line~1). 
Next, we initialize both the index $\HBIO$ and the temporary index $\mathcal{I}$ as empty sets (Line~2). 
It then iterates over the data points in increasing numerical order (Line~4). 
During each iteration, we increment $cur$ by $1$ and check if the range $[1, cur]$ is included in $R$ (Line~5). 
If yes, we update the temporary index $\mathcal{I}$ by inserting $v_i$, where $v_i$ is the data point newly included since the previous range $[1, pre]$. 
This step creates the graph index for the range $[1, cur]$.
Then, we insert $\mathcal{I}$ along with its corresponding range $[1, cur]$ into $\HBIO$ (Line~7). Finally, we update $pre$ to $cur$ for the next iteration (Line~8).
Once all points in $\mathcal{D}$ have been processed, it returns the constructed index $\HBIO$ (Line~9).

\stitle{Complexity Analysis.}
We first study the index space of $\HBIO$. 
As the graph degree is bounded by a constant $M$, the graph nodes in $\HBIO$ can be computed by the sum of range length in $R$. 
Then the summation of $[1,N/2^i]$ can be bounded by $2N$ and we get that the \underline{index size} of $\HBIO$ is $O(NM)$.
Next, we analyze the \underline{indexing time}, and from Line~4 of Algorithm~\ref{alg:build-1D-index}, we know the creation of $\HBIO$ requires $O(N)$ insertion of $\HNSW$ index.
The memory snapshot and disk storage only take a tiny portion of the overall time cost in practice.
Finally, since it only requires scanning $O (\log N)$ graph indexes for a query, and the elastic factor is bounded by 0.5, then the \underline{query complexity} is still bounded with the conclusion in Theorem~\ref{thm:half-blood-search-time-complexity}.

\stitle{Extensions.}
To achieve a more flexible elastic factor, $\frac{1}{B}$, greater or less than $0.5 = \frac{1}{2}$ for tradeoff efficiency and space, we construct graphs for the ranges $[1, N/B^i]$, where $i \in [0, \log_B N]$.
Recall that a larger elastic factor, $\frac{1}{B}$, enables more efficient query processing but with more space cost. 
On the contrary, a smaller factor saves space but reduces efficiency.
Therefore, the elastic factor acts as a tunable parameter to balance space cost and query time.

\subsection{The Method for General Queries}\label{subsec:HBI-2D}
We now introduce the process for handling a general RFAKNN query. 
To facilitate this, we define the index structure, $\HBIT$.
Similar to $\HBIO$, the $\HBIT$ index is also a hierarchical structure. 
Note that the $\HBIT$ index resembles those defined in current reconstruction-based methods for RFAKNN queries, where the hierarchical structure is encoded in a segment tree.
However, the main technical contribution lies in how to accelerate index construction and leverage elastic factors to optimize the query process.

\begin{definition}
Given a database $\mathcal{D}$ of $N$ points, the $\HBIT$ index contains $\log N$ layers and is organized as a segment tree. 
The root of the tree corresponds to the range $[1, N]$, and its children are recursively defined as divisions of the range into subranges.
\end{definition}

\begin{figure}[!t]
    \centering
    \includegraphics[scale=0.6]{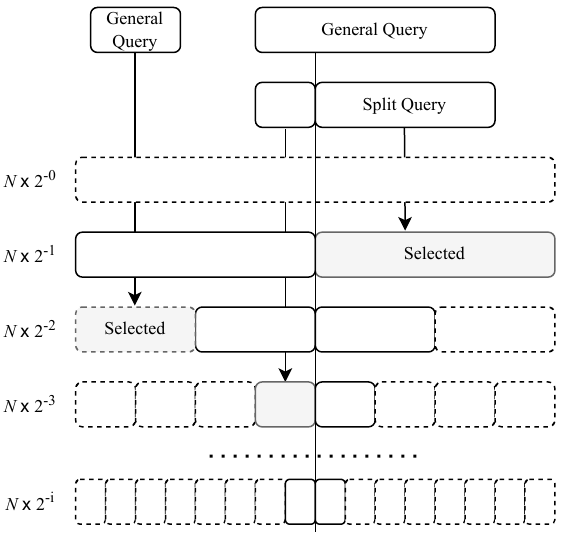}\vgap
    \vgap\caption{The Example of $\HBIT$}\vgap
    \label{fig:Query-2D-Example}
\end{figure}

Initially, a graph is constructed for the range $[1, N]$. 
This range is then recursively divided into two subranges, $[1, N/2]$ and $[N/2+1, N]$, etc. 
The resulting structure forms a segment tree, where each node corresponds to a subrange of $[1, N]$, and a graph is constructed for the associated subrange.
An example is given in Fig.~\ref{fig:Query-2D-Example}.

\begin{algorithm}[!t]
	\caption{Build-2D-Index$(\mathcal{D},[l,r])$}
	\label{alg:build-2D-index}
	\begin{small}
        \If{$r-l < \ell$}{
            \textbf{return} $\emptyset$ \tcp*{no index for small range}
        }
        $mid = \lfloor(l+r)/2\rfloor$\;
        $\mathcal{I}_l$=Build-2D-Index$(\mathcal{D},[l,mid])$\;
        $\mathcal{I}_r$=Build-2D-Index$(\mathcal{D},[mid\text{+1},r])$\;
        \If{$\mathcal{I}_l=\emptyset$}{
            create graph $\mathcal{I}_l$ on $v_i$, $i \in [l,mid]$\;
        }
        create graph $\mathcal{I}$ by inserting $v_i$ into $\mathcal{I}_l$, $i \in [mid+1, r]$\;
        $\HBIT \gets \HBIT \cup (\mathcal{I},[l,r])$\;
        \end{small}
\end{algorithm}

\begin{algorithm}[!t]
	\caption{Query-2D-Index$(\HBIT, q, [l,r], [l_q,r_q])$}
	\label{alg:Query-2D-index}
	\begin{small}
        \If{$r-l < \ell$}{
            \textbf{return} linear scan of $v_i$, $i \in [l,r]$\;
        }
        \If{$[l_q,r_q] \subseteq [l,r]$ and $\frac{|[l_q,r_q]|}{|[l,r]|} \ge c$}{
            $\mathcal{I} \gets \HBIT(l,r)$ \tcp*{select index based on range}
            \textbf{return} graph search of $q, [l,r]$ using $\mathcal{I}$
        }
        $mid = \lfloor(l+r)/2\rfloor$\;
        \tcc{search left sub-tree}
        \If{$l_q\leq mid$}{
            $S_l$=Query-2D-Index$(\HBIT,[l,mid], [l_q,\min(mid,r_q)])$\;
        }
        \tcc{search right sub-tree}
        \If{$r_q > mid$}{
            $S_r$=Query-2D-Index$(\HBIT,[mid$+1$,r],[\max(mid$+1$,l_q),r_q])$
        }
        \textbf{return} merge sort result of $S_l$ and $S_r$\;
        \end{small}
\end{algorithm}

\stitle{Indexing Algorithm.}
Algorithm~\ref{alg:build-2D-index} shows how to create $\HBIT$.  
While Algorithm~\ref{alg:build-2D-index} generates the index similarly to existing reconstruction-based methods, it is more efficient as it aims to reduce redundancy and improve performance.  
Specifically, the algorithm takes the entire range $[1, N]$ as input and calls the recursive function (Algorithm~\ref{alg:build-2D-index}) to build a segment tree.  
The recursion halts when the range becomes sufficiently small (Line~1). 
Otherwise, the current range $[l, r]$ is divided into two subranges, $[l, \text{mid}]$ and $[\text{mid}+1, r]$, and the recurision continues (Line~4-5).   
Then, if the graph index for the left subrange is not present, the algorithm incrementally inserts points $v_i$ from $i \in [l, \text{mid}]$ to construct the graph index for that range (Line~6-7).  
The most notable aspect of the approach is that instead of directly creating the graph index for the range $[l, r]$, it leverages the graph already constructed for the left subrange (created in Line~4 or Line~7) and incrementally adds points from the right subrange $[\text{mid}+1, r]$.  
This strategy avoids building the graph index from scratch, significantly enhancing efficiency.

\stitle{Query Algorithm.}
Algorithm~\ref{alg:Query-2D-index} shows how to use $\HBIT$ to process a query with point $q$ and range $[l_q, r_q]$. 
It follows a recursive process, with the termination condition being when the selected subrange $[l,r]$ becomes smaller than a predefined threshold. 
At this point, a linear scan is performed to report the answer (Line~1-2).
If the query range is $c$ times larger than the selected subrange $[l, r]$ (where $c$ is the elastic factor), the graph stored in $\HBIT$ is utilized along with $\POST$ to find the answer (Line~3-5). 
Otherwise, the subrange (initially set to $[1, N]$) is divided into two parts (Line~6), and the algorithm is applied recursively to each part: the left part (Line~7-8) and the right part (Line~9-10).
The results from the left and right parts are then merged to produce the final answer (Line~11). 
Algorithm~\ref{alg:Query-2D-index} is similar to existing reconstruction-based methods. 
Yet, a key distinction is that when a subrange encountered during traversal does not exactly match the current query range, it can still be used if Line~3 satisfies, thanks to the use of the elastic factor.

\stitle{Query Complexity.}
One might wonder whether the recursion continues until reaching the leaf node of the segment tree. Yet, the number of graph indexes that $\HBIT$ needs to select is limited to $2$.

\begin{lem}\label{lem:worse-2-index-select}
    Given an RFAKNN query, Algorithm~\ref{alg:Query-2D-index} selects at most two graph indexes in the worst case, assuming a fanout of 2.
\end{lem}

Lemma~\ref{lem:worse-2-index-select} highlights the advantages of our new method for the RFAKNN query. Specifically, if the query range is fully contained within a superset range recorded in the segment tree, and the elastic factor constraint is satisfied, the graph index corresponding to the current superset range can be directly utilized for the $\POST$ search. 
This requires only a single graph index. 
In cases where the query range is not fully contained, it is split just once. 
As a result, the query complexity remains effectively bounded by Theorem~\ref{thm:half-blood-search-time-complexity}.

\begin{example}
In Fig.~\ref{fig:Query-2D-Example}, when a query with the range $[l, r]$ is received, and it is fully contained within an indexed range $[a, b]$, we directly select $[a, b]$ if the size of $[l, r]$ is no smaller than that of $[a, b]$, adjusted by the elastic factor of 0.5. 
In contrast, the state-of-the-art method requires finding subranges that exactly cover $[l, r]$.
In cases where the above condition is not satisfied, we split the range $[l, r]$ into two parts, ensuring that each part falls within a recorded range $[a, b]$ that adheres to the elasticity factor of 0.5.
\end{example}

\stitle{Index Cost Analysis.}
We begin by analyzing the space cost of $\HBIT$. 
Given that the graph degree is bounded by a constant $M$, the graph nodes in $\HBIT$ can be computed as the sum of the range lengths in $R$. 
Since the ranges in $R$ are organized using a segment tree with $O(\log N)$ layers, and each layer consists of $N$ points, the total range length in $R$ is bounded by $N \log N$. 
Thus, the \underline{index size} of $\HBIT$ is $O(NM\log N)$.
Next, we examine the \underline{indexing time}. 
Based on the space cost analysis, the total number of nodes in $\HBIT$ is bounded by $O(N\log N)$. 
Algorithm~\ref{alg:build-2D-index} requires $O(N\log N)$ insertions into the $\HNSW$ index. 
Note that, in practice, the number of insertions in Algorithm~\ref{alg:build-2D-index} is approximately half the total number of nodes in $\HBIT$. 
This reduction occurs because the incremental construction process utilizes the left subtree, allowing each range with a left subtree to save half of the insertion time.

\begin{figure}[!t]
    \centering
    \includegraphics[scale=0.6]{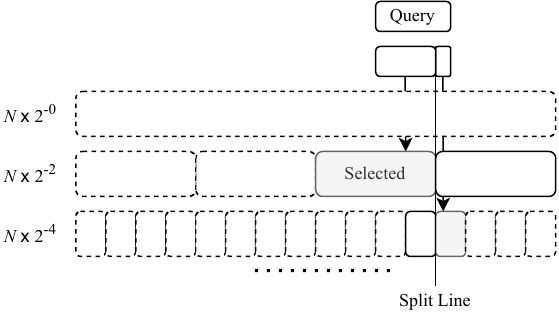}
    \vgap\caption{The Examples of a Larger Fanout}\vgap
    \label{fig:HBI-2D-Large-Fanout}
\end{figure}

\stitle{Extensions.}
In prior discussions, we fixed the fanout to $2$ and set the elastic factor to $0.5$. 
Then, we explore the extensions of a segment tree with a larger fanout and analyze their impact on both space and time complexity.
Increasing the fanout, for instance, to $4$, results in reductions in both indexing time and index space. This improvement arises as the original $\log_2(N)$ layers of the tree are reduced to $\log_4(N)$ layers, halving the costs associated with index construction in terms of both time and space. 
These improvements are illustrated in Fig.~\ref{fig:HBI-2D-Large-Fanout}.
To ensure consistency with our previous findings (see Lemma~\ref{lem:worse-2-index-select}), we observed that setting the elastic factor constraint to the reciprocal of the fanout ensures that the time complexity remains bounded. This strategy guarantees that, in the worst case, only two indexes are required.

\begin{lem}\label{lem:worse-f-index-select}
     Let $f$ denote the fanout of the segment tree and $1/f$ as the elastic factor constraint, then Algorithm~\ref{alg:Query-2D-index} also selects two graph indexes in the worst case.
\end{lem}

\begin{exmp}
    Figure~\ref{fig:HBI-2D-Large-Fanout} shows $\HBIT$ with a fanout of 4 and an elastic factor constraint of 1/4. 
    The uniform range split ensures that the query range either meets the 1/4 elastic factor constraint or spans only two sub-tree index ranges. 
    If the query range requires three or more subranges for coverage, the middle index range is fully enclosed within the query range, thereby satisfying the elastic factor constraint. 
    As a result, the original general query can still be transformed into two half-bounded queries.
\end{exmp}

Regarding the impact on query complexity, let $f$ be the fanout of the segment tree, which is a constant. The elastic factor is constrained to be $\ge 1/f$, where the probability of success (i.e., finding an in-range point) is $1/f$.
From Theorem~\ref{thm:half-blood-search-time-complexity}, the expected number of search steps is given by $E[W_k] = k \times \frac{N + 1}{Nf + 1}$, where $Nf$ denotes the number of \textit{in-range} points. 
For cases where $f$ is $1/4$ or $1/8$, the additional search steps increase by a factor of 4 and 8, respectively.
However, under the assumption that $f$ remains a fixed constant and $f < N$, the extra steps are still bounded by a constant factor of $k$. 
Consequently, the overall time complexity remains unaffected.

\section{Experiments}\label{sec:Experiments}
\subsection{Experiment Settings}
\stitle{Datasets.}
We use publicly available datasets that are widely used as benchmarks for RFAKNN search: SIFT, GLOVE, WIT, and DEEP100M. 
Among these, DEEP100M is a large-scale dataset comprising 100 million instances sampled from DEEP1B\footnote{https://research.yandex.com/blog/benchmarks-for-billion-scale-similarity-search}, which we use to evaluate scalability. 
The WIT dataset\footnote{https://github.com/google-research-datasets/wit}, which includes 2048-dimensional ResNet-50 embeddings of images from Wikipedia, was further processed by using the size of each image as the attribute value. 
All datasets used in our experiments are stored in the float32 format. 
For datasets that do not provide attribute values, we synthesize these attributes following the methods described in~\cite{SeRF-SIGMOD-2024, iRangeGraph-SIGMOD-2025}.
Details of all datasets, including the number of data points ($\mathcal{N}$), dimensionality, and query size, are summarized in Table~\ref{tab:dataset_details}. 
To form query ranges, we randomly selected two values from the range $[1, N]$ as the left and right bounds of the query range for general RFAKNN queries. For half-bounded queries, only one value was selected from $[1, N]$. We refer to these randomly generated queries collectively as ``range = mix''.
Moreover, to evaluate the impact of query range lengths on performance, we also vary range lengths as $2^{-1} \times N$, $2^{-3} \times N$, and $2^{-8} \times N$.

\stitle{Metrics.}
To evaluate the accuracy of our method, we use recall as the metric due to its extensive use in benchmarks~\cite{ann-benchmakrs,Graph-ANNS-Survey-VLDB-2021-mengzhao}. 
For the efficiency evaluation, we use the queries-per-second (QPS), which can be regarded as the number of queries that the algorithm can process in one second. 
All metrics used in the experiment are reported on averages over the query set.

\stitle{Algorithms.}
The algorithms compared in our study are as follows:

\noindent
$\bullet$
$\HBIO$: Our proposed method for Half-Bound RFAKNN queries.

\noindent
$\bullet$
$\HBIT$: Our proposed method for General RFAKNN queries.

\noindent
$\bullet$
$\SERFO$: Compression-based method for Half-Bound queries~\cite{SeRF-SIGMOD-2024}.

\noindent
$\bullet$
$\SERFT$: Compression-based method for General queries in~\cite{SeRF-SIGMOD-2024}.

\noindent
$\bullet$
$\SEG$: Reconstruction-based method in~\cite{Window-Filter-ICML-2024}.

\noindent
$\bullet$
$\SUP$: The $\SUPER$ approach~\cite{Window-Filter-ICML-2024}.

\noindent
$\bullet$
$\IRANG$: Another reconstruction-based method in~\cite{iRangeGraph-SIGMOD-2025}.

\begin{table}[t!] 
\centering 
\caption{The Statistics of Datasets} \vgap\vgap
\label{tab:dataset_details} 
\begin{footnotesize}
\begin{tabular}{c|c c c c} 
\toprule
\textbf{Dataset}& \textbf{Dimension} & \textbf{Size} & \textbf{Query Size} & \textbf{Type} \\ 
\midrule
SIFT & 128 & 1,000,000 & 1000 & Image + Attribute\\
DEEP & 96 & 1,000,000 & 1000 & Image + Attribute \\ 
GLOVE & 100 & 1,000,000 & 1000 &  Text + Attribute\\
WIT & 2048 & 1,000,000 & 1000 & Image + Attribute\\ 
DEEP100M & 96 & 100,000,000 & 1000 & Image + Attribute \\ 
\bottomrule
\end{tabular} 
\end{footnotesize}\vgap
\end{table}

\stitle{Implementation Details.}
All code was written in C++ and compiled using GCC version 9.4.0 with -Ofast optimization. 
The experiments were conducted on a workstation equipped with Intel(R) Xeon(R) Platinum 8352V CPUs @ 2.10GHz, 1TB of memory, and running Ubuntu Linux. 
We utilized multi-threading for index construction and a single thread for search evaluation.  
Since both $\IRANG$ and $\SERF$ are based on $\HNSW$, we used $\HNSW$ as the graph index for $\HBIO$, $\HBIT$, $\SEG$, and $\SUP$ to isolate and focus on the algorithmic performance.  
The default fanout for $\HBIT$ was set to 2, and we also evaluated the impact of larger fanout values in the sequel. 
The parameters of all existing algorithms were configured according to the default settings. For the algorithm using $\HNSW$ we set $M$=16, \textit{efconstruct}=200.

\input{figure/half-result-mixed}
\input{figure/general_result}
\begin{figure}[!t]
\centering
\begin{small}
\begin{tikzpicture}
    \begin{customlegend}[legend columns=4,
        legend entries={$\SERFO$,$\HBIO$,$\SERFT$,$\HBIT$},
        legend style={at={(0.45,1.15)},anchor=north,draw=none,font=\scriptsize,column sep=0.1cm}]
    \addlegendimage{line width=0.15mm,color=orange,mark=halfcircle,mark size=0.5mm}
    \addlegendimage{line width=0.15mm,color=navy,mark=otimes,mark size=0.5mm}
    \addlegendimage{line width=0.15mm,color=amber,mark=pentagon,mark size=0.5mm}
    \addlegendimage{line width=0.15mm,color=violate,mark=square,mark size=0.5mm}
    \end{customlegend}
\end{tikzpicture}\vspace{-0.5mm}
\subfloat[DEEP100M (Half)]{\vgap
\begin{tikzpicture}[scale=1]
\begin{axis}[
 height=\columnwidth/2.5,
width=\columnwidth/1.8,
xlabel=recall(\%),
ylabel=Qpsx100,
title={range=$5\cdot10^6$},
title style={yshift=-2.5mm},
label style={font=\scriptsize},
title style={font=\scriptsize},
tick label style={font=\scriptsize},
scaled ticks=false,
ymajorgrids=true,
xmajorgrids=true,
grid style=dashed,
]

\addplot[line width=0.15mm,color=orange,mark=halfcircle,mark size=0.5mm]
plot coordinates {
(88.5, 26.3502)
(93.5, 15.0904)
(94.2, 17.7897)
(96.5, 10.7117)
(97.9, 6.53141)
(98.4, 4.634)
};
\addplot[line width=0.15mm,color=navy,mark=otimes,mark size=0.5mm]
plot coordinates {
( 92.0, 73.378 )
( 96.3, 44.792 )
( 98.6, 24.994 )
( 99.5, 13.506 )
( 99.9, 7.215 )
( 100.0, 2.036 )
};

\end{axis}
\end{tikzpicture}\hspace{0.5mm}
}
\subfloat[DEEP100M (General)]{\vgap
\begin{tikzpicture}[scale=1]
\begin{axis}[
 height=\columnwidth/2.5,
width=\columnwidth/1.8,
xlabel=recall(\%),
ylabel=Qpsx1000,
title={range=$5\cdot10^6$},
title style={yshift=-2.5mm},
label style={font=\scriptsize},
title style={font=\scriptsize},
tick label style={font=\scriptsize},
scaled ticks=false,
ymajorgrids=true,
xmajorgrids=true,
grid style=dashed,
]
\addplot[line width=0.15mm,color=amber,mark=pentagon,mark size=0.5mm]
plot coordinates {
(92,10.2434)
(95, 6.49489)
(96.2, 3.84532)
(97.2, 3.39602)
};
\addplot[line width=0.15mm,color=violate,mark=square,mark size=0.5mm]
plot coordinates {
    ( 97.7, 12.245 )
    ( 98.9, 7.454 )
    ( 99.4, 4.55 )
    ( 99.9, 2.846 )
    ( 100.0, 0.649 )
};

\end{axis}
\end{tikzpicture}\hspace{0.5mm}
}

\vgap\caption{The Test of Scalability}\vgap \label{fig:large-results}

\end{small}
\end{figure}
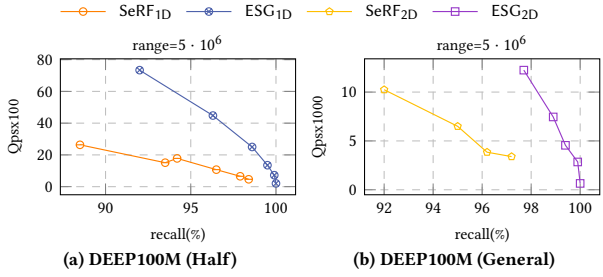
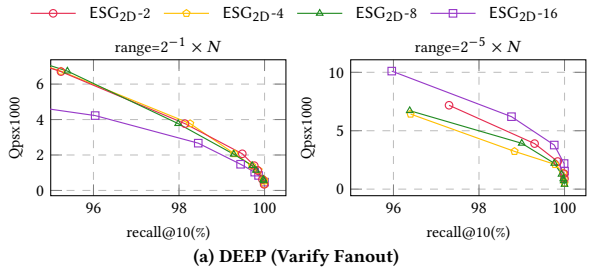
\begin{figure}[t]
\centering
\begin{footnotesize}
\begin{tikzpicture}
    \begin{customlegend}[legend columns=4,
        legend entries={$\HBIT$-2,$\HBIT$-4,$\HBIT$-8,$\HBIT$-16},
        legend style={at={(0.45,1.15)},anchor=north,draw=none,font=\scriptsize,column sep=0.1cm}]
    \addlegendimage{line width=0.15mm,color=amaranth,mark=o,mark size=0.5mm}
    \addlegendimage{line width=0.15mm,color=amber,mark=pentagon,mark size=0.5mm}
    \addlegendimage{line width=0.15mm,color=forestgreen,mark=triangle,mark size=0.5mm}
    \addlegendimage{line width=0.15mm,color=violate,mark=square,mark size=0.5mm}
    \end{customlegend}
\end{tikzpicture}
\\[-\lineskip]

\subfloat[DEEP (Varify Fanout)]{\vgap
\begin{tikzpicture}[scale=1]
\begin{axis}[
    height=\columnwidth/2.5,
width=\columnwidth/1.80,
xlabel=recall@10(\%),
ylabel=Qpsx1000,
title={range=$2^{-1}\times N$},
title style={yshift=-2.5mm},
label style={font=\scriptsize},
tick label style={font=\scriptsize},
title style={font=\scriptsize},
xmin=95,
xmax=100.5,
ymajorgrids=true,
xmajorgrids=true,
grid style=dashed,
]
\addplot[line width=0.15mm,color=violate,mark=square,mark size=0.5mm]
plot coordinates {
    ( 90.55, 6.117 )
    ( 96.04, 4.224 )
    ( 98.45, 2.667 )
    ( 99.44, 1.477 )
    ( 99.77, 1.033 )
    ( 99.86, 0.847 )
    ( 100.0, 0.467 )
};
\addplot[line width=0.15mm,color=amber,mark=pentagon,mark size=0.5mm]
plot coordinates {
    ( 88.93, 11.701 )
    ( 95.24, 6.73 )
    ( 98.26, 3.772 )
    ( 99.29, 2.079 )
    ( 99.71, 1.395 )
    ( 99.86, 1.143 )
    ( 99.98, 0.623 )
    ( 99.99, 0.545 )
    ( 100.0, 0.344 )
};
\addplot[line width=0.15mm,color=amaranth,mark=o,mark size=0.5mm]
plot coordinates {
    ( 88.92, 11.689 )
    ( 95.24, 6.704 )
    ( 98.14, 3.768 )
    ( 99.48, 2.068 )
    ( 99.76, 1.398 )
    ( 99.84, 1.124 )
    ( 99.95, 0.623 )
    ( 99.99, 0.343 )
};
\addplot[line width=0.15mm,color=forestgreen,mark=triangle,mark size=0.5mm]
plot coordinates {
    ( 88.92, 11.722 )
    ( 95.39, 6.734 )
    ( 97.98, 3.771 )
    ( 99.28, 2.043 )
    ( 99.71, 1.4 )
    ( 99.8, 1.135 )
    ( 99.97, 0.625 )
    ( 99.98, 0.544 )
};

\end{axis}
\end{tikzpicture}\hspace{0.5mm}
\begin{tikzpicture}[scale=1]
\begin{axis}[
    height=\columnwidth/2.5,
width=\columnwidth/1.80,
xlabel=recall@10(\%),
ylabel=Qpsx1000,
title={range=$2^{-5} \times N$},
title style={yshift=-2.5mm},
label style={font=\scriptsize},
tick label style={font=\scriptsize},
title style={font=\scriptsize},
xmin=95,
xmax=100.5,
ymajorgrids=true,
xmajorgrids=true,
grid style=dashed,
]
\addplot[line width=0.15mm,color=violate,mark=square,mark size=0.5mm]
plot coordinates {
    ( 95.96, 10.104 )
    ( 98.76, 6.203 )
    ( 99.76, 3.773 )
    ( 99.99, 2.192 )
    ( 100.0, 1.529 )
};
\addplot[line width=0.15mm,color=amber,mark=pentagon,mark size=0.5mm]
plot coordinates {
    ( 96.4, 6.427 )
    ( 98.83, 3.233 )
    ( 99.78, 2.115 )
    ( 99.97, 1.378 )
    ( 99.99, 0.975 )
    ( 100.0, 0.521 )
};
\addplot[line width=0.15mm,color=amaranth,mark=o,mark size=0.5mm]
plot coordinates {
    ( 97.3, 7.179 )
    ( 99.3, 3.898 )
    ( 99.84, 2.363 )
    ( 99.99, 1.322 )
    ( 100.0, 0.902 )
};
\addplot[line width=0.15mm,color=forestgreen,mark=triangle,mark size=0.5mm]
plot coordinates {
    ( 96.38, 6.717 )
    ( 99.0, 3.925 )
    ( 99.76, 2.211 )
    ( 99.93, 1.274 )
    ( 99.97, 0.883 )
    ( 99.98, 0.715 )
    ( 100.0, 0.408 )
};

\end{axis}
\end{tikzpicture}\hspace{0.5mm}
}

\vgap\caption{The Test of a Large Fanout}\label{fig:fanout-exp}
\end{footnotesize}
\end{figure}

\subsection{Experiment Results}

\stitle{Exp-1: Half-Bounded Query Performance.}
We begin by evaluating the performance of different algorithms on processing half-bounded RFAKNN queries. 
For this analysis, $\SERFO$ and $\HBIO$, both of which require only $O(N)$ space complexity, are used as comparison methods. 
Our method, $\HBIO$, is compared against $\SERFO$ across randomly generated query ranges (denoted as range = mix).
The results, shown in Fig.~\ref{fig:half-range-filter-results}, where the top-right region indicates better performance, lead to the following observations:

\noindent
$\bullet$
Our $\HBIO$ algorithm consistently outperforms $\SERFO$ across nearly all datasets, achieving average search efficiency improvements of 1.2x to 2x. This superior performance can be attributed to two main factors:  1) The integration of well-optimized search libraries within the $\HBIO$ framework enhances overall efficiency.  2)  The hierarchical structure and memory layout of $\HNSW$ are preserved in $\HBIO$, which not only minimizes engineering development costs but also significantly boosts search performance.

\begin{table}[!t]
\begin{footnotesize}
    \centering
    \caption{The Comparison of Index Time (s)}\label{tab:index-time}\vgap\vgap
    \begin{tabular}{l|cccccc}
        \toprule
        Dataset & $\SERFO$ & $\HBIO$ & $\SERFT$ & $\HBIT$ & $\SUP$ & $\IRANG$ \\ 
        \midrule
        SIFT      & 11   & 14  & 18   & 54  & 142 (2.6x)  & 475 (8.8x) \\
        DEEP      & 9    & 12  & 17   & 60 & 138 (2.3x) & 463 (7.7x) \\ 
        GLOVE     & 7    & 12  & 14   & 80  & 332 (4.2x) & 431 (5.4x) \\ 
        WIT       & 84   & 129  & 163  & 297 & 1129 (3.8x) & 2,250 (7.6x) \\ 
        DEEP100M  & 1,518 & 2,007  & 3,910 & 10,554 & -  & >24h \\ 
        \bottomrule
    \end{tabular}
\end{footnotesize}\vgap
\end{table}

\begin{table}[!t]
\begin{footnotesize}
    \centering
    \caption{The Comparison of Index Size (MB)}\label{tab:index-size}\vgap\vgap
    \begin{tabular}{l|cccccc}
        \toprule
        Dataset  & Raw Data & $\SERFT$ & $\HBIT$ &  $\SUP$   & $\IRANG$ \\ 
        \midrule
        SIFT     & 489     & 176       & 1,416    & 2,549      & 878 \\
        DEEP     & 367     & 216       & 1,416    & 2,549      & 957 \\ 
        GLOVE    & 386     & 129       & 1,416    & 2,549      & 664 \\ 
        WIT      & 7,812   & 154       & 1,416    & 2,549      & 761 \\ 
        DEEP100M & 36,621  & 22,012    & 226,630  & -          & -   \\ 
        \bottomrule
    \end{tabular}
\end{footnotesize}
\end{table}

\stitle{Exp-2: General Query Performance.}
We further investigate the performance of various methods for general RFAKNN queries. 
To evaluate their performance, we designed range queries of varying lengths, spanning from $2^{-1}$ to $2^{-8}$ of $N$. 
The primary methods compared include $\HBIT$, $\SEG$, $\SERFT$, $\SUP$, and $\IRANG$. 
Notably, the $\SEG$ algorithm utilizes the same index as $\HBIT$ but employs a different query algorithm. 
Based on the results in Fig~\ref{fig:general-range-filter-results}, we observe the following findings:

\noindent
$\bullet$
$\HBIT$ achieves comparable search efficiency to $\SUP$ at high recall levels across all datasets. Specifically, $\HBIT$ outperforms $\SUP$ on the GLOVE and WIT datasets and demonstrates strong competitiveness with $\SUP$ on the SIFT and DEEP datasets.

\noindent
$\bullet$
Our $\HBIT$ outperforms $\IRANG$ on the SIFT, DEEP, and GLOVE datasets, achieving up to a 2x improvement in query efficiency in scenarios with smaller range filters and high recall levels. 
However, we observe some performance gaps relative to $\IRANG$ on the WIT dataset and in scenarios with ranges with large lengths. 
This discrepancy stems from the need for our $\HBIT$ to compute distances to certain points outside the range. 
A potential solution to address this limitation is to incorporate distance computation acceleration techniques~\cite{PEOs-ICML-2024-kejing, Rabitq-SIGMOD-2024, ADSampling:journals/sigmod/GaoL23, Finger-WWW-2023,DDC-BSA-ICDE-2025-mingyu,MRQ-arxiv-2024}, which will be our future research.

\stitle{Exp-3: Index Time and Space.}
We compare the index time and space for various methods in Table~\ref{tab:index-time}. 
First, we observe that the indexing time of our $\HBI$ is comparable to that of $\SERF$, whereas $\IRANG$ requires 5x to 10x more time compared to $\HBI$, and $\SUP$ demands 2.7x to 4x the indexing time.  
Two primary factors contribute to this efficiency: 1) Our method leverages a well-optimized algorithm library, which ensures better concurrency and computational efficiency. 
2) It capitalizes on redundant information during the construction process by employing incremental construction based on the left subtree index, thereby further reducing the overall construction time.
Next, we examine the index space. As shown in Table~\ref{tab:index-size}, our method achieves an index size comparable to $\SERFT$ and $\IRANG$, while demonstrating a smaller space cost compared to $\SUP$.
Notably, $\SUP$ requires nearly double the space of $\HBIT$, particularly in scenarios involving lower-dimensional data.

\stitle{Exp-4: Test of Scalability.}
We further evaluate the scalability of different methods. 
To this end, we conduct experiments on the largest dataset, DEEP100M, varying the recall rate to compare different methods.
As shown in Fig.~\ref{fig:large-results}, SERFO shows a 6x lower performance compared to our proposed $\HBIO$ at 98\% recall. 
Also, $\HBIT$ maintains stable performance on large-scale datasets, whereas $\SERFT$ fails to achieve the target recall of 98\%.
Our method also offers significant advantages in space efficiency and indexing time on large-scale datasets by controlling the fanout size.
For instance, with a fanout of 16 (the efficiency result in appendix), $\HBIT$-16 requires only 70GB of storage---just one-third of the space required for a fanout of 2---and 5830 seconds of construction time.

\stitle{Exp-5: Test of Top-1 Search.}
We compare different methods of processing range-filtering approximate nearest neighbor search (where $k=1$). 
As shown in Fig.~\ref{fig:general-range-filter-results}(a) and (b), our proposed method, $\HBIT$, consistently achieves a 1.3x to 2x improvement compared to the $\IRANG$ and $\SUP$. 
Our theoretical analysis reveals that the number of steps required to identify the global nearest neighbor scales sublinearly with the dataset size, while the additional search steps increase linearly with $k$. 
This provides a dramatic advantage for our method in RFANN search.

\stitle{Exp-6: Test of Large Fanout.}
We set the fanout to $2$ for the segment tree in our method, $\HBIT$. 
Next, we test $\HBIT$ when it is configured with a larger fanout. 
As shown in Fig.~10, a larger fanout provides the benefit of more cheap index space. 
For instance, on the DEEP dataset, $\HBIT$-4, $\HBIT$-8, and $\HBIT$-16—corresponding to fanouts of 4, 8, and 16---require only 709MB, 567MB, and 426MB of space, respectively. 
We further analyze search efficiency in Fig.~11. 
The results indicate that $\HBIT$-2 achieves relatively good overall query efficiency. 
However, it does not exhibit a big performance advantage compared to configurations with larger fanouts, which is consistent with our theoretical analysis.
Additionally, $\HBIT$-16 even shows some advantages for smaller range filters. 
This is primarily due to the more relaxed elastic factor constraint (>1/16), which increases the likelihood of using a single index to efficiently answer general RFANN search queries.

\section{Related Work}\label{sec:Related}
\stitle{Attribute-filtering AKNN Search.}
Attribute-filtering approximate $k$ Nearest Neighbor (AFAKNN) search represents a more general case of RFAKNN search. While RFAKNN focuses solely on a single numeric attribute, AFAKNN incorporates additional attribute values, such as discrete labels, to broaden its applicability.
Existing studies~\cite{AnalyticDB-VLDB-2020,Vbase-OSDI-2023,CAPS-arxiv-2023,NHQ-NIPS-2022-mengzhao-wang} address AFAKNN search from both database systems and algorithmic perspectives~\cite{Window-Filter-ICML-2024}. 
1) From the database systems perspective, query cost prediction is a core strategy for allocating appropriate search methods to AFAKNN queries. Systems such as ADB~\cite{AnalyticDB-VLDB-2020}, Vbase~\cite{Vbase-OSDI-2023}, VSAG~\cite{VSAG-arxiv-Mingyu}, and Milvus~\cite{Milvus-SIGMOD-2021} exemplify this approach. 
2) On the algorithmic side, several methods aim to enhance query efficiency through attribute-specific optimizations. Filtered-DiskANN~\cite{Filtered-diskann-WWW-2023} improves efficiency by predicting attribute values, NHQ~\cite{NHQ-NIPS-2022-mengzhao-wang} refines the graph-index structure based on attribute values, and HQI~\cite{HQI-SIGMOD-2023} processes queries offline by leveraging the workload. The recent work UNG~\cite{UNG-SIGMOD-2025} proposes a unified approach to AFAKNN search by constructing a label navigating graph for attribute values.
Despite these advancements, there remains a significant performance gap between AFAKNN algorithms and specialized RFAKNN search algorithms, such as $\SERF$, $\IRANG$, and $\SUPER$~\cite{SeRF-SIGMOD-2024,Window-Filter-ICML-2024,iRangeGraph-SIGMOD-2025}. As such, our baseline evaluation focuses exclusively on highly competitive algorithms like $\SERF$, $\IRANG$, and $\SUPER$ to ensure robust and meaningful comparisons.

\stitle{Optimizing Segment Tree.}
Current reconstruction-based methods for RFAKNN search queries primarily rely on segment trees, which require $O(\log N)$ subranges during query processing. 
To reduce the number of required subranges, the $\SUPER$ algorithm proposed in~\cite{Window-Filter-ICML-2024} redundantly stores certain ranges within the index. 
However, this approach comes with a significant trade-off: it incurs nearly double the space overhead compared to partitioning with segment trees.
As an empirical analysis, our proposed $\HBIT$ method is competitive with $\SUPER$ and demonstrates superior performance in top-1 search (when $k=1$ for RFAKNN search). 
In addition, our work provides a theoretical analysis of the $\POST$ algorithm, establishing that at most $2$ subranges are required during query processing. 
This robust theoretical foundation highlights the novelty of our approach in addressing RFAKNN search queries effectively.

\stitle{Similarity Search in General Spaces.}
In addition to Euclidean distance, other distance metrics such as inner product and angular distance are widely used in high-dimensional AKNN search.
When the inner product is selected as the distance metric, it forms the problem of Maximum Inner Product Search (MIPS), which can be efficiently addressed using LSH-based methods~\cite{ALSH(MIPS)-NIPS-2014,FARGO-VLDB-2023,RangLSH-NIPS-2018} with theoretical guarantees. 
Notably, graph-based indexes are also extensively applied to MIPS~\cite{Understanding-Graph(MIPS)-AAAI-2020,Non-metirc-Graph(MIPS)-NIPS-2018}, often demonstrating superior search performance compared to LSH-based approaches.
In this paper, we propose methods based on the $\POST$ principle within Euclidean space, accompanied by rigorous theoretical analysis. 
The core idea is to filter out range-restricted k-nearest neighbors (KNN) from the global KNN search.
We also aim to extend this approach to handle scenarios where the inner product is used as the distance metric, enabling solutions for more general distance or similarity search problems.

\section{Conclusions}\label{sec:Conclusions}
We focus on the problem of RFAKNN queries by introducing a paradigm shift from the commonly used $\PRE$ principle to the $\POST$ principle. 
This transition enables a theoretical analysis within a more generalized query complexity framework for $\POST$, particularly when incorporating the elastic factor.
Building upon this theoretical foundation, we propose novel reconstruction-based methods that guarantee at most two subranges are required for a query, compared to the $O(\log N)$ subranges employed by existing approaches. 
Extensive experimental results demonstrate the superiority of our methods in query efficiency while preserving query accuracy. 
In future work, we aim to extend this concept to other index structures, focusing on enhancing construction efficiency and optimizing space utilization.

\newpage
\balance
\bibliographystyle{ACM-Reference-Format}
\bibliography{sample-base}

\newpage

\end{document}